\documentclass[useAMS,usenatbib]{mnras}
\usepackage{graphicx}
\usepackage{amsmath}
\usepackage{color}
\usepackage{float}
\usepackage{caption}
\usepackage{placeins}
\usepackage{lipsum}
\usepackage{multicol}

\usepackage{amssymb}

\def\be{\begin{equation}}
\def\ee{\end{equation}}

\def\gsim{\lower.5ex\hbox{\gtsima}}
\def\lsim{\lower.5ex\hbox{\ltsima}} \def\gtsima{$\; \buildrel > \over \sim \;$} \def\ltsima{$\; \buildrel < \over \sim \;$} \def\prosima{$\;
\buildrel \propto \over \sim \;$} \def\gsim{\lower.5ex\hbox{\gtsima}}
\def\lsim{\lower.5ex\hbox{\ltsima}}
\def\simgt{\lower.5ex\hbox{\gtsima}}
\def\simlt{\lower.5ex\hbox{\ltsima}}
\def\simpr{\lower.5ex\hbox{\prosima}}  
 
 \def\gtsima{$\; \buildrel > \over \sim \;$}
\def\ltsima{$\; \buildrel < \over \sim \;$}
\def\gsim{\lower.5ex\hbox{\gtsima}}
\def\lsim{\lower.5ex\hbox{\ltsima}}
\def\simgt{\lower.5ex\hbox{\gtsima}}
\def\simlt{\lower.5ex\hbox{\ltsima}}
\def\simpr{\lower.5ex\hbox{\prosima}}

\def\E3{{\cal E}_{\rm g}^{III}}

\def\Msun{\rm M_\odot}

\def\Msun{\rm M_\odot}

\def\M*{M_*}
\def\Z*{Z_*}
\def\L*{L_*}

\def\lnofb{{\rm L^{nofb}_{1375}}}
\def\lfb{{\rm L^{uvfb}_{1375}}}
\def\muv{{\rm M_{UV}}}

\newcommand{\bear}{\begin{eqnarray}}
\newcommand{\ear}{\end{eqnarray}}
\newcommand{\nline}{\notag \\} 
\newcommand{\f}{\frac}

\newcommand{\de}{\ensuremath{{\rm d}}}

\newcommand{\fig}[1]{Figure~\ref{#1}}

\title[HFF constraints on fluctuating UVB]{Probing the fluctuating Ultra-violet background using the Hubble Frontier Fields}
\author[Choudhury \& Dayal]{Tirthankar Roy Choudhury$^{1}$\thanks{tirth@ncra.tifr.res.in} \& Pratika Dayal$^2$ \\
$^{1}$ National Centre for Radio Astrophysics, Tata Institute of Fundamental Research, Pune 411007, India\\
$^{2}$ Kapteyn Astronomical Institute, University of Groningen, P.O. Box 800, 9700 AV Groningen, The Netherlands\\
}
\begin{document}

\date{}

\maketitle

\begin{abstract}
In recent years, the rise in the number of Lyman Break Galaxies detected at high redshifts $z \geq 6$ has opened up the possibility of understanding early galaxy formation physics in great detail. In particular, the faint-end slope ($\alpha$) of the Ultra-violet luminosity function (UV LF) of these galaxies is a potential probe of feedback effects that suppress star formation in low mass haloes. In this work, we propose a {\it proof-of-concept} calculation for constraining the fluctuating UV background during reionization by constraining $\alpha$ in different volumes of the Universe. Because of patchy reionization, different volumes will experience different amount of photo-heating which should lead to a scatter in the measured $\alpha$. Our approach is based on a simple model of the UV LF that is a scaled version of the halo mass function combined with an exponential suppression in the galaxy luminosity at the faint-end because of UV feedback. Although current data is not sufficient to constrain $\alpha$ in different fields, we expect that, in the near future, observations of the six lensed Hubble Frontier Fields with the James Webb Space Telescope ({\it JWST}) will offer an ideal test of our concept.
\end{abstract}


\section{Introduction}
The past few years have seen an enormous increase in the observational data collected for galaxies that had formed in the first billion years of the Universe thanks to a combination of state of the art observatories (most notably the Hubble Space Telescope; {\it HST}) as well as refined selection methods. In the latter category, the Lyman Break technique has been exceptionally successful at building up a statistically significant repository of $z \gsim 6$ Lyman Break Galaxies \citep[LBGs; e.g.][]{mclure2009,mclure2010,mclure2013,livermore2017,bouwens2015,
bouwens2010a,bowler2014a,atek2015,oesch2014}. The measured ultra-violet (UV) luminosity (between $1250-1500$\AA\, in the rest-frame) from the above-mentioned works has been been used to construct the evolving UV luminosity function (UV LF) all the way to $z \sim 10$ allowing unprecedented studies on the key feedback physics of early galaxies. One of the key feedback effects is associated with Type II supernovae that can potentially heat or blow-out a significant (or even all) of the gas content in low-mass halos \citep[e.g.][]{1999ApJ...513..142M}. The second feedback effect is that associated with cosmic reionization in the redshift range $15 \gtrsim z \gtrsim 6$ \citep{fan2006, stark2011, planck2018}.

During reionization, photoionization heating from the continually rising UV background (UVB) can raise the gas temperature to about $2 \times 10^4$ K in ionized regions \citep{miralda-escude1994}, which, in principle, could result in the UVB photo-evaporating gas from the lowest mass galaxies suppressing further star formation. Given that many existing models assume these galaxies to be the key reionization sources \citep{choudhury2007,finlator2011,wise2014,robertson2015,2017ApJ...836...16D}, the impact of this UV feedback is critical both for galaxy formation as well as the process of reionization.

However, so far, the fluctuating UVB has only been measured at relatively low-redshifts \citep[$z \sim 5-6$;][]{2015MNRAS.447.3402B,2015MNRAS.453.2943C,2017MNRAS.465.3429C}. Further, since the baryonic content of a halo exposed to a UVB depends on a multitude of parameters, including the redshift, the thermal history and the intensity of the UVB, the halo baryon fraction during reionization remains a matter of debate \citep{okamoto2008, wise2012b, hasegawa2013, sobacchi2013b}. A number of works find the lowest mass haloes to be impervious to the UVB unless the key reionization sources are either molecular-cooling driven \citep{sobacchi2013b} rapidly losing their gas after SN explosions \citep{pawlik2015} or low-mass galaxies that contain little/no molecular gas in the first place \citep{gnedin2014}. On the other hand, other works find the UVB to suppress the star formation rate at high-$z$ \citep{petkova2011, finlator2011, hasegawa2013}. Naturally while the first school of thought would predict no impact of the UVB on the UV LF \citep[e.g.][]{gnedin2014}, in the latter case, the faint-end slope of the UV LF (typically denoted by $\alpha$) would become shallower due to the decreasing star formation efficiencies of low-mass haloes \citep[see e.g.][]{2015ApJ...806...67D,bremer2018}.

In this paper, we propose a {\it proof-of-concept} calculation that uses the observations of the faint-end of the UV LF in different fields to yield hints on the fluctuating UVB. Our calculations are based on the premise that supernova feedback, effectively depending on the ratio between the star formation rate and halo potential should be the same in every field observed, barring cosmic variance. On the other hand, feedback from a fluctuating UVB can potentially result in UV LF faint-end slopes that will vary from field to field. This is an ideal time to undertake such analyses given that the James Webb Space Telescope ({\it JWST}) is expected to re-observe the six lensed Hubble Frontier Fields yielding a significant sample of $z \gsim 6$ galaxies extending to UV magnitudes as faint as $\muv \sim -12.5$.

\begin{figure*}
\center{\includegraphics[width=0.75\textwidth]{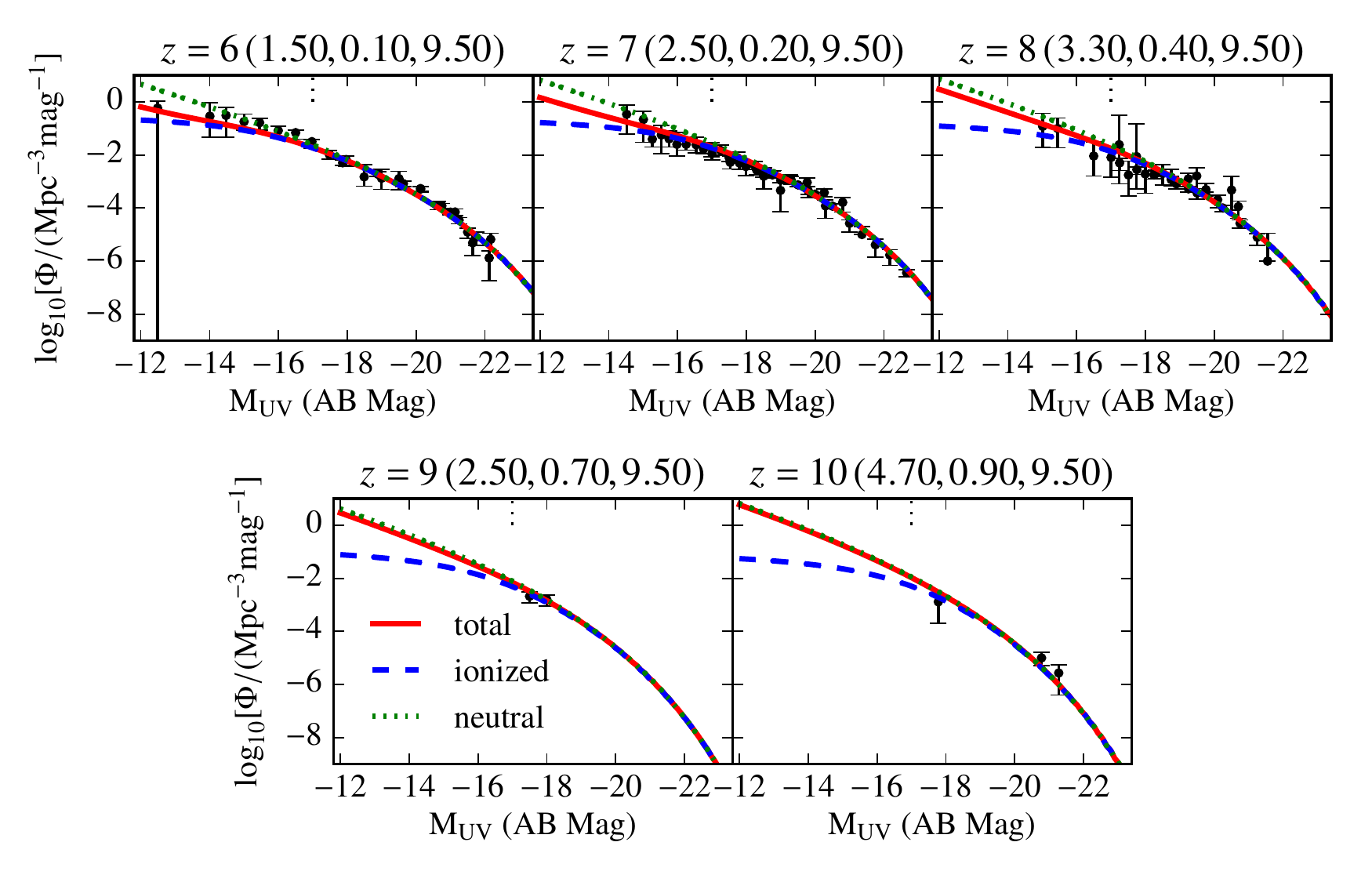}}
\caption{The evolving UV luminosity function (LF) for $z \simeq 6-10$ with the model parameter values [$10^3 \epsilon_*, Q_{\rm HI}, \log_{10}(M_{\rm crit} / \Msun)]$ as marked at the top of each panel. The points with error-bars represent the observational data
\citep{mclure2009,livermore2017, bouwens2015,
bouwens2010a, mclure2010, mclure2013, bowler2014a,
atek2015,
oesch2014}, while the different curves show the predictions from our model. The green dotted (blue dashed) curves are the UV LFs for the neutral (ionized) regions. Note that the faint end of the LFs in the ionized regions are affected by UV feedback. The red solid curves denote the globally averaged UV LF. }
\label{fig:lumfun_z}
\end{figure*}

\section{Theoretical model}

\subsection{Modelling the Ultra-violet luminosity function}

The modelling of galaxy formation, in general, involves a number of complex physical processes \citep[for reviews on different aspects of galaxy formation, see, e.g.][]{1988RvMP...60....1O,2005ARA&A..43..769V,2007ARA&A..45..565M,2014ARA&A..52..291C,2015arXiv151103457K,2015ARA&A..53...51S}. The simplest models assume that each dark matter halo contains only one galaxy and the luminosity of the galaxy is primarily determined by the corresponding halo mass. In that case, the observed UV LF can be modelled as a scaled halo mass function (HMF) at that redshift.

In this work, we assume that in absence of any feedback, the UV luminosity of a halo is proportional to the halo mass, $M_h$, such that
\be
\lnofb(M_h) = \epsilon_* ~\left(\f{\Omega_b}{\Omega_m}\right) ~ M_h~ l_{1375},
\label{eq:L_Mh}
\ee
where the term $(\Omega_b/\Omega_m)$ represents the cosmological baryon fraction. Further, $l_{1375}=10^{33.07}~\mbox{erg~s}^{-1}$ \AA$^{-1}~\Msun^{-1}$ is the specific ultra-violet luminosity for a newly formed stellar population assuming a metallicity of $5\%$ of the solar value and a Salpeter initial mass function (IMF) between $0.1-100\Msun$. Finally, $\epsilon_*$ is the fraction of baryons in the halo that get converted into stars. Physically, $\epsilon_*$ is the product of the baryon fraction that can cool and the cold gas fraction that can form stars. We assume the combination $\epsilon_*~l_{1375}$ to be independent of $M_h$ (although it can depend on $z$). Note that any deviation of $l_{1375}$ from this fiducial value can be absorbed within the unknown parameter $\epsilon_*$.

The relation between $l_{1375}$ and $M_h$ gets modified in presence of feedback processes. The radiative feedback arising from the UVB can suppresses the gas fraction in low mass haloes in ionized regions. We assume that the decrease in the total galaxy luminosity due to this UV radiative feedback can be modelled through the simple relation \citep[e.g.][]{sobacchi2013b}
\be
\lfb(M_h) = \epsilon_* ~2^{- M_{\rm crit} / M_h}~\left(\f{\Omega_b}{\Omega_m}\right)~M_h~l_{1375},
\label{eq:L_Mh_fb}
\ee
where $M_{\rm crit}$ is the critical halo mass characterizing the effect of feedback. In fact, the above form implies that the luminosity of a galaxy in a halo of mass $M_{\rm crit}$ ($0.1 M_{\rm crit}$) decreases by a factor $2$ ($\sim 1000$) in presence of feedback. Although more complicated forms for UV feedback suppression exist in the literature \citep{2000ApJ...542..535G}, the above simple form has been shown to serve the purpose of modelling the evolving UV LF at high redshift \citep[see e.g.][]{2015ApJ...806...67D}.

The UV luminosities obtained above can be converted to an absolute UV magnitude (in the standard AB system) using $\muv = -2.5 \log_{10}({\rm L}_{1375}) + 51.60$ where ${\rm L}_{1375}$ is the total UV luminosity (in ${\rm erg\, s^{-1} \, {Hz}^{-1}}$) from the galaxy.

Naturally, the UVB will be non-zero only in volumes that are ionized, while neutral regions would be devoid of any ionizing photons. Consequently, radiative feedback will suppress the gas content in only those galaxies which form in already ionized regions. If $Q_{\rm HI}$ is the \emph{neutral} volume fraction of the universe, we expect that a fraction $Q_{\rm HII} \equiv (1 - Q_{\rm HI})$ of galaxies will be affected by feedback \citep{2005MNRAS.361..577C,2017ApJ...836...16D}. Under these assumptions, one can compute the globally averaged  UV LF as a combination of a fully-suppressed UV LF in ionized regions ($\Phi^{\rm uvfb}$) and an unaffected UV LF ($\Phi^{\rm nofb}$) in neutral regions such that
\bear
\Phi(\muv) \!\!\!\!\!\! &=& \!\!\!\!\!\! (1 - Q_{\rm HI})~\Phi^{\rm uvfb}(\muv) + Q_{\rm HI}~\Phi^{\rm nofb}(\muv)
\nline
\!\!\!\!\!\! &=& \!\!\!\!\!\! \f{\de n}{\de M_h} \left[Q_{\rm HII}~\f{\de M_h}{\de \lfb}~\f{\de \lfb}{\de \muv}
+ Q_{\rm HI}~\f{\de M_h}{\de \lnofb}~\f{\de \lnofb}{\de \muv} \right],
\nline
\ear
where $\de n / \de M_h$ is the halo mass function\footnote{In this work, we use the HMF ($\de n / \de M_h$) of \citet{1999MNRAS.308..119S,2001MNRAS.323....1S}. We use a flat $\Lambda$CDM cosmology with $\Omega_m = 0.308, \Omega_b = 0.0482, h = 0.678, n_s = 0.961, \sigma_8 = 0.829$ \citep{2014A&A...571A..16P}.}.
Thus in our model the UV LF can be calculated once we fix three parameters: $\epsilon_*, M_{\rm crit}$ and $Q_{\rm HI}$.

\begin{table}

\begin{center}

\begin{tabular}{|c|ccccc|}
\hline
$z$ & 6 & 7 & 8 & 9 & 10 \\
\hline
$10^3 \epsilon_*$ & 1.5 & 2.5 & 3.3 & 2.5 & 4.7 \\
\hline

\end{tabular}

\end{center}

\caption{Values of $\epsilon_*$ constrained from the bright-end of the UV LF at the different redshifts shown in Columns 2-6.}
\label{tab:eps_*}
\end{table}

\subsubsection{Constraints on the star formation efficiency}
\label{sfe}
We start by discussing the observational constraints on the star formation efficiency parameter $\epsilon_*$. When $M_h \gg M_{\rm crit}$, the haloes hosting galaxies are so massive that UV feedback effects are quite unimportant and in that case, the UV LF becomes independent of $Q_{\rm HI}$ and is entirely determined by the single free parameter $\epsilon_*$. We can exploit the above fact and fix the value of $\epsilon_*$ by comparing our predicted UV LF with the observations at the bright end (${\rm M_{UV}} \lsim -17$) as shown (by green dotted lines) in Fig. \ref{fig:lumfun_z}. The values of $\epsilon_*$ obtained by this comparison are listed in Table~\ref{tab:eps_*} at each $z$ .

We also show the feedback affected UV LF appropriate for galaxies in the feedback-affected HII regions (by blue dashed lines in the same figure). In order to compute these, we fix the value of $M_{\rm crit} = 10^{9.5} \Msun$ independent of the redshift which is consistent with the findings of, e.g., \citet{2000ApJ...542..535G}. For each redshift, we choose the value of the the third free parameter $Q_{\rm HI}$ so that the total UV LF (red solid lines in the same figure) gives a reasonable visual fit to the available data. The respective values of the 3 free parameters, [$10^3 \epsilon_*, Q_{\rm HI}, \log_{10} (M_{\rm crit} / \Msun)$], are indicated above each panel of the figure. This essentially shows that there exist combinations of the three parameters which can provide a satisfactory fit to the data for this simplified model of the evolving UV LF. The effect of UV feedback, as one can see from the figure, is to essentially flatten the faint-end slope of the UV LF which is a direct consequence of the suppression of luminosity in low-mass galaxies. It is worth mentioning that the currently available data points at the faint-end are not accurate enough to constrain $M_{\rm crit}$ and $Q_{\rm HI}$ stringently because of their large error-bars -- it is therefore quite possible that there exist other combinations of the parameter values which can provide an equally good fit to the data.

\begin{figure*}
\center{\includegraphics[trim=0 0 0 10,clip,width=0.85\textwidth]{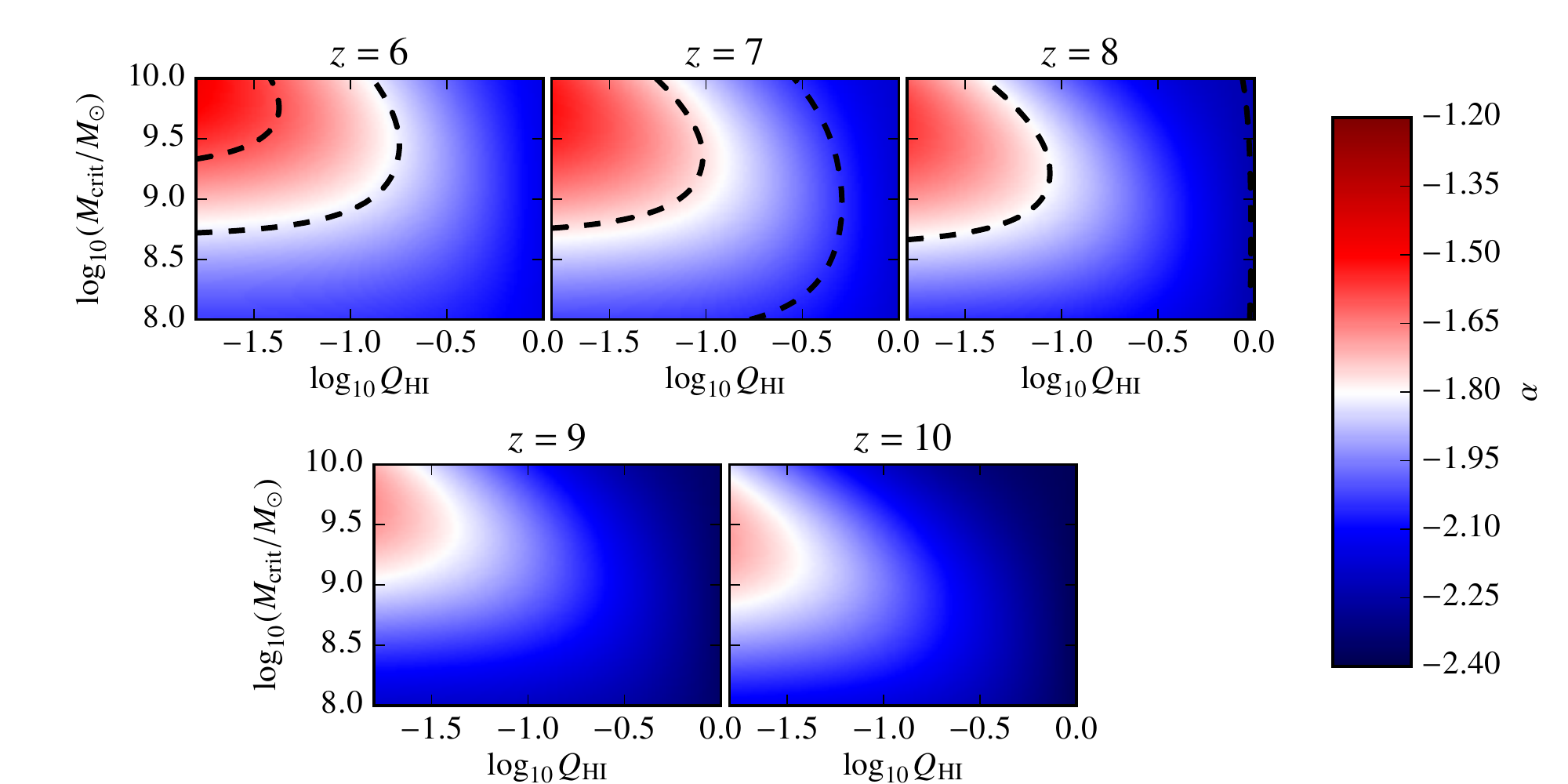}}
\caption{The dependence of the faint-end slope $\alpha$ of the UV LF (corresponding to the red solid curves in \fig{fig:lumfun_z}) on $M_{\rm crit}$ and $Q_{\rm HI}$ for different redshifts. The black dashed curves in the three panels in the top row denote the allowed $1-\sigma$ ranges in $\alpha$ obtained from the available observational data.}
\label{fig:faint_end_slope}
\end{figure*}

\subsubsection{Constraints on the fluctuating UVB}

We now extend the concepts described in the previous section to probe the impact of UV feedback from a patchy ionizing background. Given that UV feedback directly only affects the faint-end slope, we now restrict our discussions to constraining the value of $\alpha$ using observations from forthcoming facilities such as the {\it JWST}. For definiteness, we define the faint end as consisting of galaxies with ${\rm M_{UV}} \gsim -17$, although minor variations of this threshold are not expected to affect our conclusions.

Since the parameter $\epsilon_*$ (Sec. \ref{sfe} above) is already fixed by the bright-end, we can compute $\alpha$ for all possible combinations of $M_{\rm crit}$ and $Q_{\rm HI}$. The plot of $\alpha$ as a function of $M_{\rm crit}$ and $Q_{\rm HI}$ is shown in \fig{fig:faint_end_slope}. To understand the dependence of $\alpha$ on the two parameters, let us concentrate on the first panel on the left hand side ($z = 6$). When the universe is mostly neutral $Q_{\rm HI} \to 1$, UV feedback effects are quite negligible resulting in $\alpha$ being independent of $M_{\rm crit}$. At the other extreme, when $Q_{\rm HI} \to 0$, we find that the slope flattens ($\alpha$ increases) with increasing $M_{\rm crit}$ (for a fixed $Q_{\rm HI}$). This is simply because UV feedback becomes more severe and hence leads to suppression in the luminosity from an increasing fraction of low-mass haloes. For a fixed value of the critical halo mass, say, $M_{\rm crit} \sim 10^{9} - 10^{10} \Msun$, we find that the slope flattens with decreasing $Q_{\rm HI}$. This effect arises because of UV feedback affecting a larger fraction of $M_h \lsim M_{\rm crit}$ haloes. Interestingly, we find that the slope is largely independent of $Q_{\rm HI}$ for $M_{\rm crit} \sim 10^{8} - 10^{8.5} \Msun$. This is because for such small values of the critical mass, UV feedback only affects the lowest-mass galaxies which are below the observational limits. The same qualitative conclusions hold for the other redshifts as well. We find that for the same value of $M_{\rm crit}$ and $Q_{\rm HI}$, the slope is steeper at higher redshifts. This is because the HMF at the small mass end steepens with increasing redshift.

We also show in the figure the presently available observational constraints on $\alpha$ taken from \citet{2014MNRAS.445.2545D}. The two dashed lines in each panel show the $1-\sigma$ limits at the corresponding redshift. Interestingly, one can constrain $Q_{\rm HI} < 0.2 (0.5)$ at $z = 6 (7)$ at $1-\sigma$ confidence level with the available data.  Clearly the constraints degrade as we go to higher redshifts because of the lack of data points at the faint-end and hence it is almost impossible to put any constraint on $\alpha$ at $z \geq 8$.

Although the effect of the radiative feedback on the UV LF has been well-studied \citep[see, e.g.,][]{2007MNRAS.377..285S,2014NewA...30...89S,2016MNRAS.463.1968Y,2017MNRAS.464.1633F,2018arXiv180505945S}, the discussion above provides a rather quantitative and direct way to constrain UV feedback parameters using the observed UV LF. However, the underlying model used suffers from a significant shortcoming which is related to the degeneracies between different types of feedback. E.g., (type II) supernova feedback would also tend to suppress star formation in low and intermediate mass haloes, and can potentially lead to flattening in the faint-end slope \citep{1999ApJ...513..142M,2003MNRAS.339..312S,2007ApJ...670....1G,2012MNRAS.421.3522H,2014NewA...30...89S}. While one can, in principle, incorporate the effects of SN feedback in the model we are using, this would lead to more free parameters and it would become almost impractical to constrain the parameters with sufficient accuracy. This then warrants the question whether observations of flat $\alpha$ do indeed allow us to probe the patchy UV background in presence of other complicated physical processes. This degeneracy between different feedbacks affecting the faint-end of the UV LF can, in principle, be lifted by observing different volumes or fields on the sky. If the process of reionization is indeed patchy, as is predicted by almost all existing models, it is expected that the ionization and thermal states of the intergalactic medium (IGM) in different volumes would be different. In that case, the UVB and the impact of UV feedback (for galaxies having the same luminosity) would vary from field to field which would be manifested as a scatter in $\alpha$. It is worth emphasising that supernova feedback, which depends on the balance between the star formation rate and the underlying dark matter halo potential, is not expected to change from field to field (except for the cosmic variance). We thus propose that one can study the effects of radiative feedback by observing the UV LF across a number of different fields.

Once we measure the value of $\alpha$ to sufficient accuracy in different patches of the sky, we can use the panels of \fig{fig:faint_end_slope} to put constraints of $M_{\rm crit}$ and $Q_{\rm HI}$ for \emph{each patch}, assuming that we have already fixed $\epsilon_*$ using the bright-end. Assuming that $M_{\rm crit}$ does not vary across fields, this would allow us to constrain $Q_{\rm HI}$ in each field. Any scatter in $\alpha$ and hence $Q_{\rm HI}$ would allow us to constrain the UVB fluctuations. As is clear, it is not possible to obtain sufficiently constrained values of $\alpha$ in individual field with the current data. However, in the very near future, the {\it JWST} is expected to re-observe the six lensed Hubble Frontier Fields. Given its capability of observing down to $\muv \sim -15$, combined with moderate lensing magnifications of a factor of 10, we expect a significant sample of $z \gsim 6$ galaxies extending to magnitudes as faint as $\muv \sim -12.5$ over $\sim 10\times 10$ Mpc patches. The scatter in the value of $\alpha$ from these fields would provide an ideal test of patchy UV feedback at high-$z$ using the faint-end of the UV LF.

\section{Summary}

In recent times, the availability of high-quality data on high redshift Lyman Break Galaxies (LBGs), particularly the UV luminosity function (UV LF), has opened up the possibility of understanding various physical processes related to early galaxy formation in great detail. 
We present a proof-of-concept calculation based on the faint-end of the UV LF to constrain the fluctuating UV background (UVB) during reionization. As per our current understanding, the photo-heating arising from UV radiation will suppress star formation in low mass haloes in ionized regions. Because 
reionization is patchy, the severity of this feedback will be different in different volumes of the universe. With this in mind, our concept consists of (i) a simple model of UV LF based on scaled halo mass function, combined with an exponential suppression of the star formation in galaxies formed in ionized regions, and (ii) comparing the model with the observed UV LF in different patches in the sky. The scatter in the UV LF across different patches, in principle, should probe the patchy UV feedback at high redshifts. The currently available data is not sensitive enough to constrain the fluctuating UVB by measuring the LF in different patches of the sky. One expects that, in the very near future, the {\it JWST} will re-observe the six lensed Hubble Frontier Fields with unprecedented sensitivity, thus enabling measurement of the faint-end slope of the UV LF in different patches. These observations would serve as ideal tests of our proof-of-concept.

Finally we comment on possible complications to be accounted for while comparing the model with the data. Firstly, 
in addition to the patchy UVB, there could be some scatter in the UV LF across different patches arising from the underlying cosmic variance. Furthermore, the clustering of galaxies would lead to correlation between their positions and the feedback-affected ionized regions. All such issues are best addressed through numerical simulations, which we plan to take up in more detail in the future.

\section*{Acknowledgments}
TRC acknowledges support from the Associateship Scheme of ICTP, Trieste. PD acknowledges support from the European Research Council's starting grant DELPHI (717001) and from the European Commission's and University of Groningen's CO-FUND Rosalind Franklin program. PD thanks R. Bouwens, N. Gnedin, P. Oesch and Z. Haiman for illuminating discussions.

\bibliographystyle{mnras}
\bibliography{mybib,fb}

\begin{thebibliography}{51}
\expandafter\ifx\csname natexlab\endcsname\relax\def\natexlab#1{#1}\fi

\bibitem[{{Atek} {et~al}\mbox{.}(2015){Atek}, {Richard}, {Jauzac}, {Kneib},
  {Natarajan}, {Limousin}, {Schaerer}, {Jullo}, {Ebeling}, {Egami}, \&
  {Clement}}]{atek2015}
{Atek} H. {et~al.}, 2015, \apj, 814, 69

\bibitem[{{Becker} {et~al}\mbox{.}(2015){Becker}, {Bolton}, {Madau}, {Pettini},
  {Ryan-Weber}, \& {Venemans}}]{2015MNRAS.447.3402B}
{Becker} G.~D., {Bolton} J.~S., {Madau} P., {Pettini} M., {Ryan-Weber} E.~V.,
  {Venemans} B.~P., 2015, \mnras, 447, 3402

\bibitem[{{Bouwens} {et~al}\mbox{.}(2010){Bouwens}, {Illingworth},
  {Gonz{\'a}lez}, {Labb{\'e}}, {Franx}, {Conselice}, {Blakeslee}, {van Dokkum},
  {Holden}, {Magee}, {Marchesini}, \& {Zheng}}]{bouwens2010a}
{Bouwens} R.~J. {et~al.}, 2010, \apj, 725, 1587

\bibitem[{{Bouwens} {et~al}\mbox{.}(2015){Bouwens}, {Illingworth}, {Oesch},
  {Trenti}, {Labb{\'e}}, {Bradley}, {Carollo}, {van Dokkum}, {Gonzalez},
  {Holwerda}, {Franx}, {Spitler}, {Smit}, \& {Magee}}]{bouwens2015}
{Bouwens} R.~J. {et~al.}, 2015, \apj, 803, 34

\bibitem[{{Bowler} {et~al}\mbox{.}(2014){Bowler}, {Dunlop}, {McLure}, {Rogers},
  {McCracken}, {Milvang-Jensen}, {Furusawa}, {Fynbo}, {Taniguchi}, {Afonso},
  {Bremer}, \& {Le F{\`e}vre}}]{bowler2014a}
{Bowler} R.~A.~A. {et~al.}, 2014, \mnras, 440, 2810

\bibitem[{{Bremer} {et~al}\mbox{.}(2018){Bremer}, {Dayal}, \&
  {Ryan-Weber}}]{bremer2018}
{Bremer} J., {Dayal} P., {Ryan-Weber} E.~V., 2018, \mnras

\bibitem[{{Chardin} {et~al}\mbox{.}(2015){Chardin}, {Haehnelt}, {Aubert}, \&
  {Puchwein}}]{2015MNRAS.453.2943C}
{Chardin} J., {Haehnelt} M.~G., {Aubert} D., {Puchwein} E., 2015, \mnras, 453,
  2943

\bibitem[{{Chardin} {et~al}\mbox{.}(2017){Chardin}, {Puchwein}, \&
  {Haehnelt}}]{2017MNRAS.465.3429C}
{Chardin} J., {Puchwein} E., {Haehnelt} M.~G., 2017, \mnras, 465, 3429

\bibitem[{{Choudhury} \& {Ferrara}(2005)}]{2005MNRAS.361..577C}
{Choudhury} T.~R., {Ferrara} A., 2005, \mnras, 361, 577

\bibitem[{{Choudhury} \& {Ferrara}(2007)}]{choudhury2007}
{Choudhury} T.~R., {Ferrara} A., 2007, \mnras, 380, L6

\bibitem[{{Conselice}(2014)}]{2014ARA&A..52..291C}
{Conselice} C.~J., 2014, \araa, 52, 291

\bibitem[{{Dayal} {et~al}\mbox{.}(2017){Dayal}, {Choudhury}, {Bromm}, \&
  {Pacucci}}]{2017ApJ...836...16D}
{Dayal} P., {Choudhury} T.~R., {Bromm} V., {Pacucci} F., 2017, \apj, 836, 16

\bibitem[{{Dayal} {et~al}\mbox{.}(2014){Dayal}, {Ferrara}, {Dunlop}, \&
  {Pacucci}}]{2014MNRAS.445.2545D}
{Dayal} P., {Ferrara} A., {Dunlop} J.~S., {Pacucci} F., 2014, \mnras, 445, 2545

\bibitem[{{Dayal} {et~al}\mbox{.}(2015){Dayal}, {Mesinger}, \&
  {Pacucci}}]{2015ApJ...806...67D}
{Dayal} P., {Mesinger} A., {Pacucci} F., 2015, \apj, 806, 67

\bibitem[{{Fan} {et~al}\mbox{.}(2006){Fan}, {Carilli}, \& {Keating}}]{fan2006}
{Fan} X., {Carilli} C.~L., {Keating} B., 2006, \araa, 44, 415

\bibitem[{{Finlator} {et~al}\mbox{.}(2011){Finlator}, {Dav{\'e}}, \&
  {{\"O}zel}}]{finlator2011}
{Finlator} K., {Dav{\'e}} R., {{\"O}zel} F., 2011, \apj, 743, 169

\bibitem[{{Finlator} {et~al}\mbox{.}(2017){Finlator}, {Prescott},
  {Oppenheimer}, {Dav{\'e}}, {Zackrisson}, {Livermore}, {Finkelstein},
  {Thompson}, \& {Huang}}]{2017MNRAS.464.1633F}
{Finlator} K. {et~al.}, 2017, \mnras, 464, 1633

\bibitem[{{Gnedin}(2000)}]{2000ApJ...542..535G}
{Gnedin} N.~Y., 2000, \apj, 542, 535

\bibitem[{{Gnedin} \& {Kaurov}(2014)}]{gnedin2014}
{Gnedin} N.~Y., {Kaurov} A.~A., 2014, \apj, 793, 30

\bibitem[{{Greif} {et~al}\mbox{.}(2007){Greif}, {Johnson}, {Bromm}, \&
  {Klessen}}]{2007ApJ...670....1G}
{Greif} T.~H., {Johnson} J.~L., {Bromm} V., {Klessen} R.~S., 2007, \apj, 670, 1

\bibitem[{{Hasegawa} \& {Semelin}(2013)}]{hasegawa2013}
{Hasegawa} K., {Semelin} B., 2013, \mnras, 428, 154

\bibitem[{{Hopkins} {et~al}\mbox{.}(2012){Hopkins}, {Quataert}, \&
  {Murray}}]{2012MNRAS.421.3522H}
{Hopkins} P.~F., {Quataert} E., {Murray} N., 2012, \mnras, 421, 3522

\bibitem[{{Krumholz}(2015)}]{2015arXiv151103457K}
{Krumholz} M.~R., 2015, ArXiv e-prints

\bibitem[{{Livermore} {et~al}\mbox{.}(2017){Livermore}, {Finkelstein}, \&
  {Lotz}}]{livermore2017}
{Livermore} R.~C., {Finkelstein} S.~L., {Lotz} J.~M., 2017, \apj, 835, 113

\bibitem[{{Mac Low} \& {Ferrara}(1999)}]{1999ApJ...513..142M}
{Mac Low} M.-M., {Ferrara} A., 1999, \apj, 513, 142

\bibitem[{{McKee} \& {Ostriker}(2007)}]{2007ARA&A..45..565M}
{McKee} C.~F., {Ostriker} E.~C., 2007, \araa, 45, 565

\bibitem[{{McLure} {et~al}\mbox{.}(2009){McLure}, {Cirasuolo}, {Dunlop},
  {Foucaud}, \& {Almaini}}]{mclure2009}
{McLure} R.~J., {Cirasuolo} M., {Dunlop} J.~S., {Foucaud} S., {Almaini} O.,
  2009, \mnras, 395, 2196

\bibitem[{{McLure} {et~al}\mbox{.}(2013){McLure}, {Dunlop}, {Bowler},
  {Curtis-Lake}, {Schenker}, {Ellis}, {Robertson}, {Koekemoer}, {Rogers},
  {Ono}, {Ouchi}, {Charlot}, {Wild}, {Stark}, {Furlanetto}, {Cirasuolo}, \&
  {Targett}}]{mclure2013}
{McLure} R.~J. {et~al.}, 2013, \mnras, 432, 2696

\bibitem[{{McLure} {et~al}\mbox{.}(2010){McLure}, {Dunlop}, {Cirasuolo},
  {Koekemoer}, {Sabbi}, {Stark}, {Targett}, \& {Ellis}}]{mclure2010}
{McLure} R.~J., {Dunlop} J.~S., {Cirasuolo} M., {Koekemoer} A.~M., {Sabbi} E.,
  {Stark} D.~P., {Targett} T.~A., {Ellis} R.~S., 2010, \mnras, 403, 960

\bibitem[{{Miralda-Escud{\'e}} \& {Rees}(1994)}]{miralda-escude1994}
{Miralda-Escud{\'e}} J., {Rees} M.~J., 1994, \mnras, 266, 343

\bibitem[{{Oesch} {et~al}\mbox{.}(2014){Oesch}, {Bouwens}, {Illingworth}, \&
  et. al.}]{oesch2014}
{Oesch} P.~A., {Bouwens} R.~J., {Illingworth} G.~D., et. al., 2014, \apj, 786,
  108

\bibitem[{{Okamoto} {et~al}\mbox{.}(2008){Okamoto}, {Gao}, \&
  {Theuns}}]{okamoto2008}
{Okamoto} T., {Gao} L., {Theuns} T., 2008, \mnras, 390, 920

\bibitem[{{Ostriker} \& {McKee}(1988)}]{1988RvMP...60....1O}
{Ostriker} J.~P., {McKee} C.~F., 1988, Reviews of Modern Physics, 60, 1

\bibitem[{{Pawlik} {et~al}\mbox{.}(2015){Pawlik}, {Schaye}, \& {Dalla
  Vecchia}}]{pawlik2015}
{Pawlik} A.~H., {Schaye} J., {Dalla Vecchia} C., 2015, \mnras, 451, 1586

\bibitem[{{Petkova} \& {Springel}(2011)}]{petkova2011}
{Petkova} M., {Springel} V., 2011, \mnras, 412, 935

\bibitem[{{Planck Collaboration} {et~al}\mbox{.}(2014){Planck Collaboration},
  {Ade}, {Aghanim}, {Armitage-Caplan}, {Arnaud}, {Ashdown}, {Atrio-Barandela},
  {Aumont}, {Baccigalupi}, {Banday}, \& et~al.}]{2014A&A...571A..16P}
{Planck Collaboration} {et~al.}, 2014, \aap, 571, A16

\bibitem[{{Planck Collaboration} {et~al}\mbox{.}(2018){Planck Collaboration},
  {Aghanim}, {Akrami}, {Ashdown}, {Aumont}, {Baccigalupi}, {Ballardini},
  {Banday}, {Barreiro}, {Bartolo}, {Basak}, {Battye}, {Benabed}, {Bernard},
  {Bersanelli}, {Bielewicz}, {Bock}, {Bond}, {Borrill}, {Bouchet}, {Boulanger},
  {Bucher}, {Burigana}, {Butler}, {Calabrese}, {Cardoso}, {Carron},
  {Challinor}, {Chiang}, {Chluba}, {Colombo}, {Combet}, {Contreras}, {Crill},
  {Cuttaia}, {de Bernardis}, {de Zotti}, {Delabrouille}, {Delouis}, {Di
  Valentino}, {Diego}, {Dor{\'e}}, {Douspis}, {Ducout}, {Dupac}, {Dusini},
  {Efstathiou}, {Elsner}, {En{\ss}lin}, {Eriksen}, {Fantaye}, {Farhang},
  {Fergusson}, {Fernandez-Cobos}, {Finelli}, {Forastieri}, {Frailis},
  {Franceschi}, {Frolov}, {Galeotta}, {Galli}, {Ganga}, {G{\'e}nova-Santos},
  {Gerbino}, {Ghosh}, {Gonz{\'a}lez-Nuevo}, {G{\'o}rski}, {Gratton},
  {Gruppuso}, {Gudmundsson}, {Hamann}, {Handley}, {Herranz}, {Hivon}, {Huang},
  {Jaffe}, {Jones}, {Karakci}, {Keih{\"a}nen}, {Keskitalo}, {Kiiveri}, {Kim},
  {Kisner}, {Knox}, {Krachmalnicoff}, {Kunz}, {Kurki-Suonio}, {Lagache},
  {Lamarre}, {Lasenby}, {Lattanzi}, {Lawrence}, {Le Jeune}, {Lemos},
  {Lesgourgues}, {Levrier}, {Lewis}, {Liguori}, {Lilje}, {Lilley}, {Lindholm},
  {L{\'o}pez-Caniego}, {Lubin}, {Ma}, {Mac{\'{\i}}as-P{\'e}rez}, {Maggio},
  {Maino}, {Mandolesi}, {Mangilli}, {Marcos-Caballero}, {Maris}, {Martin},
  {Martinelli}, {Mart{\'{\i}}nez-Gonz{\'a}lez}, {Matarrese}, {Mauri}, {McEwen},
  {Meinhold}, {Melchiorri}, {Mennella}, {Migliaccio}, {Millea}, {Mitra},
  {Miville-Desch{\^e}nes}, {Molinari}, {Montier}, {Morgante}, {Moss}, {Natoli},
  {N{\o}rgaard-Nielsen}, {Pagano}, {Paoletti}, {Partridge}, {Patanchon},
  {Peiris}, {Perrotta}, {Pettorino}, {Piacentini}, {Polastri}, {Polenta},
  {Puget}, {Rachen}, {Reinecke}, {Remazeilles}, {Renzi}, {Rocha}, {Rosset},
  {Roudier}, {Rubi{\~n}o-Mart{\'{\i}}n}, {Ruiz-Granados}, {Salvati}, {Sandri},
  {Savelainen}, {Scott}, {Shellard}, {Sirignano}, {Sirri}, {Spencer},
  {Sunyaev}, {Suur-Uski}, {Tauber}, {Tavagnacco}, {Tenti}, {Toffolatti},
  {Tomasi}, {Trombetti}, {Valenziano}, {Valiviita}, {Van Tent}, {Vibert},
  {Vielva}, {Villa}, {Vittorio}, {Wandelt}, {Wehus}, {White}, {White},
  {Zacchei}, \& {Zonca}}]{planck2018}
{Planck Collaboration} {et~al.}, 2018, ArXiv e-prints

\bibitem[{{Robertson} {et~al}\mbox{.}(2015){Robertson}, {Ellis}, {Furlanetto},
  \& {Dunlop}}]{robertson2015}
{Robertson} B.~E., {Ellis} R.~S., {Furlanetto} S.~R., {Dunlop} J.~S., 2015,
  \apjl, 802, L19

\bibitem[{{Samui}(2014)}]{2014NewA...30...89S}
{Samui} S., 2014, \na, 30, 89

\bibitem[{{Samui} {et~al}\mbox{.}(2007){Samui}, {Srianand}, \&
  {Subramanian}}]{2007MNRAS.377..285S}
{Samui} S., {Srianand} R., {Subramanian} K., 2007, \mnras, 377, 285

\bibitem[{{Samui} {et~al}\mbox{.}(2018){Samui}, {Srianand}, \&
  {Subramanian}}]{2018arXiv180505945S}
{Samui} S., {Srianand} R., {Subramanian} K., 2018, ArXiv e-prints

\bibitem[{{Sheth} {et~al}\mbox{.}(2001){Sheth}, {Mo}, \&
  {Tormen}}]{2001MNRAS.323....1S}
{Sheth} R.~K., {Mo} H.~J., {Tormen} G., 2001, \mnras, 323, 1

\bibitem[{{Sheth} \& {Tormen}(1999)}]{1999MNRAS.308..119S}
{Sheth} R.~K., {Tormen} G., 1999, \mnras, 308, 119

\bibitem[{{Sobacchi} \& {Mesinger}(2013)}]{sobacchi2013b}
{Sobacchi} E., {Mesinger} A., 2013, \mnras, 432, 3340

\bibitem[{{Somerville} \& {Dav{\'e}}(2015)}]{2015ARA&A..53...51S}
{Somerville} R.~S., {Dav{\'e}} R., 2015, \araa, 53, 51

\bibitem[{{Springel} \& {Hernquist}(2003)}]{2003MNRAS.339..312S}
{Springel} V., {Hernquist} L., 2003, \mnras, 339, 312

\bibitem[{{Stark} {et~al}\mbox{.}(2011){Stark}, {Ellis}, \&
  {Ouchi}}]{stark2011}
{Stark} D.~P., {Ellis} R.~S., {Ouchi} M., 2011, \apjl, 728, L2

\bibitem[{{Veilleux} {et~al}\mbox{.}(2005){Veilleux}, {Cecil}, \&
  {Bland-Hawthorn}}]{2005ARA&A..43..769V}
{Veilleux} S., {Cecil} G., {Bland-Hawthorn} J., 2005, \araa, 43, 769

\bibitem[{{Wise} {et~al}\mbox{.}(2012){Wise}, {Abel}, {Turk}, {Norman}, \&
  {Smith}}]{wise2012b}
{Wise} J.~H., {Abel} T., {Turk} M.~J., {Norman} M.~L., {Smith} B.~D., 2012,
  \mnras, 427, 311

\bibitem[{{Wise} {et~al}\mbox{.}(2014){Wise}, {Demchenko}, {Halicek}, {Norman},
  {Turk}, {Abel}, \& {Smith}}]{wise2014}
{Wise} J.~H., {Demchenko} V.~G., {Halicek} M.~T., {Norman} M.~L., {Turk} M.~J.,
  {Abel} T., {Smith} B.~D., 2014, \mnras, 442, 2560

\bibitem[{{Yue} {et~al}\mbox{.}(2016){Yue}, {Ferrara}, \&
  {Xu}}]{2016MNRAS.463.1968Y}
{Yue} B., {Ferrara} A., {Xu} Y., 2016, \mnras, 463, 1968

\end{thebibliography}

\label{lastpage}
\end{document}